\documentclass[prb,twocolumn,amsmath,amssymb,showpacs]{revtex4}
\usepackage{graphicx}

\def\beq{\begin{eqnarray}}
\def\eeq{\end{eqnarray}}
\def\beqa{\begin{eqnarray}}
\def\eeqa{\end{eqnarray}}

\begin{document}

\title{Self-energy renormalization around the flux phase
in the $t-J$ model: Possible implications in underdoped
cuprates
}

\author{Andr\'es Greco}

\affiliation{
Facultad de Ciencias Exactas, Ingenier\'{\i}a y Agrimensura and Instituto de F\'{\i}sica Rosario (UNR-CONICET).
Av. Pellegrini 250-2000 Rosario-Argentina.
}

\date{\today}

\begin{abstract}
The flux phase predicted by the $t-J$ model in the large-$N$ limit  exhibits features that make it a candidate for describing the 
pseudogap regime of cuprates. However certain properties, as for instance the 
prediction of well defined quasiparticle peaks, speak against this scenario. 
We have addressed these issues by computing self-energy renormalizations in the vicinity to flux phase. 
The calculated spectral functions show  features similar to those observed in experiments. At low doping, near the flux phase, the
spectral functions  are  anisotropic on the Fermi surface and  
very incoherent near the hot spots. The temperature and doping evolution of self-energy and spectral functions are discussed
and compared with the experiment.
\end{abstract}


\maketitle

The pseudogap (PG) phase of high-$T_c$ cuprates presents unusual properties in clear contrast 
to those expected for conventional metals\cite{timusk99} and,
in spite of the many scenarios  proposed, the problem remains open.
A well known phenomenological theory is the $d$-CDW\cite{chakravarty01} in which a normal state (NS) gap  was proposed. 
Interestingly, $d$-CDW is closely related to the flux
phase (FP) which was formerly predicted in the context of the $t-J$ model.\cite{affleck88} In both scenarios the PG coexists and competes with superconductivity, is complex and  shows $d$-wave symmetry.
FP was also studied by means of different analytical and numerical
methods.\cite{morse90,cappelluti99,foussats04,leung00}
In the mean field approximation of the  
$t-J$ model,  below a characteristic temperature $T^*$, a $d$-wave  NS gap opens on the Fermi surface (FS)
near the hot spots (${\bf k}\cong (0,\pi)$). This 
leads to pockets  with low spectral 
weight intensity in the outer  section.\cite{chakravarty03} Thus, these pockets resemble the arcs observed 
in ARPES experiments.\cite{norman98}
Furthermore, recent progress on  the $t-J$ model shows that the competition between FP and superconductivity leads to a phase diagram qualitatively similar to that observed  in cuprates.\cite{cappelluti99} 
Raman and tunneling spectroscopy results show similarities with the experiment.\cite{zeyher02}
Very recent ARPES,\cite{kando07} Raman,\cite{letacon07}  specific heat,\cite{wen07} and femtosecond optical pulses\cite{chia07} experiments together with some theoretical developments\cite{theo} support the  competing two-gap scenario.
From  the above reasons FP may be considered as  a good candidate  for describing the PG 
region.
Nevertheless there are some drawbacks to be addressed. While in the FP scenario a true phase transition occurs at  $T^*$, many  experiments show  a smooth behavior as a function of temperature.\cite{tallon01}
In addition,  well defined quasiparticle (QP) peaks are predicted everywhere on the Brilloiun zone (BZ), above and below $T^*$, while experiments show 
that the QP exists only near nodal direction.\cite{kim98} Near hot spots, coherent peaks are observed neither 
below nor above the PG temperature in apparent contradiction with the FP picture.
On the other hand, the PG observed in ARPES, consistently with tunneling experiments,\cite{renner98} seems to 
be filling up but is not closing, \cite{norman1998} leading to a smooth evolution  of spectral properties.

In spite of the above objections we will assume the point of view that the basic physics of cuprates is contained in the $t-J$ model (see Ref.[\onlinecite{lee06}] for discussions). 
Moreover, to address the above discussion it is necessary to perform a controllable self-energy and spectral function calculation
beyond mean field. 
Recently it was proposed 
a large-$N$ approach for the $t-J$ model 
based on the path integral representation for $X$-operators.\cite{foussats04}
At mean field level the approach of Ref.[\onlinecite{foussats04}] agrees with slave boson\cite{affleck88,morse90} and the approach of Ref.[\onlinecite{cappelluti99}]. Additionally, it can be easily applied beyond mean field allowing the 
calculation of self-energy and spectral functions.\cite{bejas06}
In this paper we discuss spectral functions and self-energy calculations in the NS 
in the vicinity to the FP instability. We have found that the spectral functions and the 
corresponding self-energies are very anisotropic on the FS leading to well defined QP near nodal direction and very incoherent features near hot spots.

We study the two dimensional  $t-J$ model with  hopping $t$ and Heisenberg coupling $J/t=0.3$ between nearest neighbors sites 
on the square lattice.\cite{model}
In the following  energy is in units of $t$.
Using the formulation described in Ref.[\onlinecite{foussats04}] 
the mean field homogeneous Fermi liquid (HFL) becomes unstable when the static ($\omega=0$) flux susceptibility $\chi_{flux}({\bf q},\omega)=(\frac{\delta}{2})^2 [(8/J)\Delta^2-\Pi({\bf q},\omega)]^{-1}$
diverges. $\Pi({\bf q},\omega)$ 
is an electronic bubble calculated with a form factor 
$\gamma({\bf q},{\bf k})=2 \Delta (sin(k_x-q_x/2)-sin(k_y-q_y/2))$.\cite{foussats04,bejas06,flux}
(For definition of $\Delta$ see below).
In Fig.1a the phase diagram in the doping-temperature ($\delta-T$) plane is presented where  $T^*$ indicates the FP 
instability onset.  At $T=0$ a phase transition occurs 
at the critical doping $\delta_c \cong 0.12$ for the incommensurate critical vector ${\bf q}_c\cong(1,0.86)\pi$ near $(\pi,\pi)$. At finite $T$ 
the instability is commensurate (${\bf q}_c=(\pi,\pi)$).  
Therefore, since the instability takes place at, or close to, 
$(\pi,\pi)$ the form factor $\gamma({\bf q},{\bf k})$ transforms 
into $\sim (cos(k_x)-cos(k_y))$ which indicates the $d$-wave character of the FP.
Since $T^*$ 
falls abruptly to zero at $\delta_c$,\cite{tp} we associate the FP with the 
scenario discussed in Fig.1a of Ref.[\onlinecite{tallon01}].  
In Fig.1b we have plotted
$Im \chi_{flux}({\bf q}=(\pi,\pi),\omega)$ at $T=0$ which shows that when $\delta \rightarrow \delta_c$  a low energy $d$-wave 
flux  mode becomes soft accumulating large spectral weight. 
This soft mode reaches $\omega=0$ at $\delta=\delta_c$ freezing the FP.
\begin{figure}
\begin{center}
\setlength{\unitlength}{1cm}
\includegraphics[width=9cm,angle=0.]{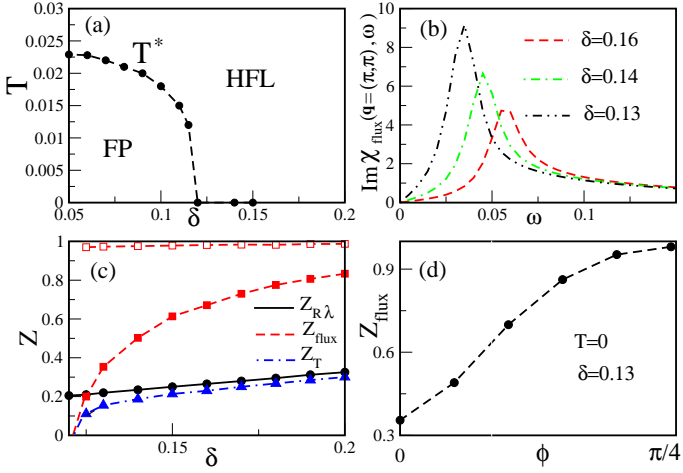}
\end{center}
\caption{(Color online)
(a) Large-$N$ $t-J$ model phase diagram in the $\delta-T$ plane. (b) Softening of the $d$-wave flux mode for several $\delta$
values approaching $\delta_c \cong 0.12$. (c) QP weight $Z$ for $\Sigma_{R \lambda}$ (black circles), $\Sigma_{flux}$ (red squares) and $\Sigma_T$ 
(blue triangles).
(d) $Z_{flux}$ versus the FS angle $\phi$ for $\delta=0.13$.}
\label{fig:fase}
\end{figure}

For studying the soft mode influence on one electron spectral properties we calculate self-energy renormalizations in the NS  in the vicinity of $T^*$.
The inclusion of the superconducting ground state in the calculation does not change significantly the FP region (see Refs.[\onlinecite{cappelluti99}] and [\onlinecite{zeyher02}]). 
After collecting all contributions in ${\cal{O}}(1/N)$ the  self-energy can be written as\cite{bejas06} (in ${\cal{O}}(1/N)$ the
inclusion of  self-energy corrections in the  calculation of electronic bubbles 
is not necessary)
\begin{eqnarray}\label{eq:SigmaIm0}
Im \, \Sigma_T({\mathbf{k}},\omega)=Im \, \Sigma_{R \lambda}({\mathbf{k}},\omega)+Im \, \Sigma_{flux}({\mathbf{k}},\omega)
\end{eqnarray}
\noindent where 
\begin{eqnarray}\label{eq:SigmaIm0}
Im \, \Sigma_{R \lambda}({\mathbf{k}},\omega)&=&-\frac{1}{ N_{s}}
\sum_{{\mathbf{q}}} \left\{ \Omega^{2} \;
 Im[D_{RR}({\mathbf{q}},\omega-\epsilon_{{k-q}})] \right. \nonumber\\
&& \hspace{-1cm} + \; 2\;\Omega \;
Im[D_{\lambda R}({\mathbf{q}},\omega-\epsilon_{{k-q}})]\nonumber\\
&& \hspace{-1cm}+ \left.
Im[D_{\lambda \lambda}({\mathbf{q}},\omega-\epsilon_{{k-q}})] \right\} \nonumber\\
&& \hspace{-1cm}\times \left[n_{F}(-\epsilon_{{k-q}}) + n_{B}(\omega-\epsilon_{{k-q}})\right]
 , \nonumber \\
\end{eqnarray}
\noindent and 
\begin{eqnarray}\label{eq:SigmaIm0}
Im \, \Sigma_{flux}({\mathbf{k}},\omega)&=&-\frac{1}{N_{s}}\sum_{{\mathbf{q}}} \gamma^2({\bf q},{\bf {k}}) Im \chi_{flux}({\bf q},\omega-\epsilon_{{k-q}}) \nonumber\\
&& \hspace{-1cm}\times \left[n_{F}(-\epsilon_{{k-q}}) + n_{B}(\omega-\epsilon_{{k-q}})\right]
\end{eqnarray}
In the above expressions, $\Omega=(\varepsilon_{{k-q}}+\omega+\mu)/2 $,  $\varepsilon_{k}=(t\delta+\Delta)
(cos(k_x)+cos(k_y))$ is the mean field electronic dispersion, $\epsilon_{k}= \varepsilon_{k}-\mu$ and,
$\Delta=\frac{J}{2N_s} \sum_k cos(k_x) n_F(\epsilon_{k})$. $n_{B}$ ($n_F$) is the Bose (Fermi) factor,  $\mu$ is the 
chemical potential and $N_s$ is the number of sites.
$\Sigma_{R \lambda}$ corresponds to 
the charge ($\delta R$) sector,  nondouble occupancy  ($\delta \lambda$) sector and the mixing of both.\cite{bejas06,2x2} 
Since for  $J=0.3$ there is no important  influence of $J$-contributions in 
$\Sigma_{R \lambda}$ then, $\Sigma_{R \lambda}$ is  close to the self-energy for $J=0$. 
Instead, $\Sigma_{flux}$ is dominated by $J$. The explicit expressions of $D_{RR}$, $D_{\lambda R}$ and $D_{\lambda \lambda}$ are given in  
Ref.[\onlinecite{bejas06}].

In Fig.1c we have plotted the QP weight $Z=1/(1-\partial Re \Sigma/\partial \omega)$ vs $\delta$ for each self-energy contribution.
$Z_{R \lambda}$ (black circles) are results at the Fermi crossing momentum in the  $(0,0)$-$(0,\pi)$ direction (antinodal-$k_F$).
$Z_{R \lambda}$ decreases (linearly)  with decreasing doping ($Z_{R \lambda} \rightarrow 0$ when $\delta  \rightarrow 0$).
Due to the fact that $\Sigma_{R\lambda}$ is very isotropic on the FS,  
results for $Z_{R \lambda}$ at the Fermi crossing momentum in the  nodal  direction (nodal-$k_F$)
are very similar to those for the antinodal-$k_F$. 
In contrast, $Z_{flux}$ (red squares) is strongly anisotropic on the FS. 
For nodal-$k_F$ (open red squares) $Z_{flux} \sim 1$  which means that $\Sigma_{flux}$ has no important contribution near 
nodal direction. For antinodal-$k_F$  (closed red squares) $Z_{flux}$ is rapidly decreasing  with doping, {\it i.e.},
$Z_{flux}\rightarrow 0$ when $\delta \rightarrow \delta_c$ where the flux instability takes place. 
In order to show explicitly the anisotropy of $\Sigma_{flux}$, in Fig.1d 
we have plotted  $Z_{flux}$ versus the FS angle $\phi$ from $\phi=0$ (antinodal-$k_F$) to 
$\phi=\pi/4$ (nodal-$k_F$), for $\delta=0.13$. 
The soft $d$-wave mode (Fig.1b), which scatters
electrons on the FS with momentum transfer ${\bf q} \sim (\pi,\pi)$,  is responsible for the observed anisotropy.
Using $Z_{R\lambda}$ and $Z_{flux}$  the total QP weight can be written as $Z_T=1/(Z_{R \lambda}^{-1}+Z_{flux}^{-1}-1)$. 
Blue triangles are the results for $Z_T$ at  the antinodal-$k_F$. 
While for large $\delta$ self-energy renormalizations are dominated by $\Sigma_{R \lambda}$, $\Sigma_{flux}$ dominates near $\delta_c$. 
Results for $Z_T$ at the nodal-$k_F$ (not shown) 
follow the same trend as $Z_{R\lambda}$, {\it i.e.}, the nodal $Z_T$ exhibits a linear behavior with slope similar to that discussed in Ref.[\onlinecite{yoshida03}].
\begin{figure}
\begin{center}
\setlength{\unitlength}{1cm}
\includegraphics[width=9cm,angle=0.]{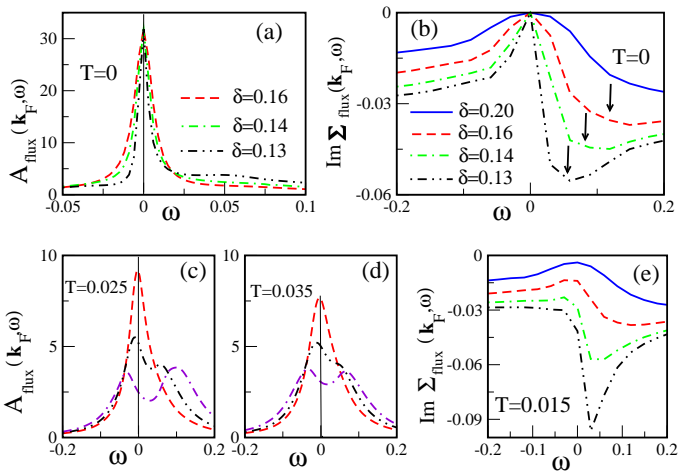}
\end{center}
\caption{(Color online)
(a) $T=0$ spectral functions at the antinodal-$k_F$ for several doping values approaching $\delta_c$. A broadening $\eta=0.01$ 
was used in the calculation. The vertical line marks the Fermi level.
(b) $T=0$ flux scattering rate for several $\delta$ values at the antinodal-$k_F$.
(c) and (d) Spectral functions at the antinodal-$k_F$ for $T=0.025$ and $T=0.035$ respectively, for $\delta=0.16$ (red dashed line), $\delta=0.13$ (black double-dotted-dashed line) and 
$\delta=0.10$ (violet double-dashed-dotted line). (e)
The same as in (b) but for $T=0.015$. 
}
\label{fig:fase}
\end{figure}

Next we will consider the $\Sigma_{flux}$ contribution to the spectral function $A_{flux}({\bf k},\omega)$.
Spectral functions for different $\delta$ values  near $\delta_c$ at the  antinodal-$k_F$   are presented for  $T=0$ in Fig.2a.
When $\delta \rightarrow \delta_c$, the area under 
the QP peak decreases (consistently with the behavior of
$Z_{flux}$) and 
spectral weight is transfered to the incoherent structure at about $\omega \sim 0.05-0.1t$. 
Fig.2b shows the $\omega$ behavior of the flux scattering rate $Im \Sigma_{flux}$ at the antinodal-$k_F$ for several $\delta$ 
values near $\delta_c$. 
With decreasing doping towards $\delta_c$, $Im \Sigma_{flux}$ evolves from a FL-like behavior ($\sim \omega^2$) 
(see solid  line 
for $\delta=0.20$) to a non FL-like behavior ($\sim \omega$) (see double-dotted-dashed line for $\delta=0.13$).
At the same time, the structures (marked with arrows) that appear at low energy are correlated with the soft mode 
discussed in Fig.1b.
$Im \Sigma_{flux}$ at the nodal-$k_F$ (not shown) is weaker  than 
for the antinodal-$k_F$  and, in addition, it shows a FL-like  behavior for all $\delta$ values. 
In summary, as observed in experiments, \cite{kaminsky05,chang07} at low doping the scattering rate is strongly anisotropic on the FS presenting a FL-like ($\sim \omega^2$) behavior near the nodal-$k_F$ and a non FL-like  ($\sim \omega$) behavior near the antinodal-$k_F$. In addition, this anisotropy disappears with overdoping where the behavior is more FL-like.

In Figs.2c and 2d spectral functions at the antinodal-$k_F$  are presented for $T=0.025$ 
and $T=0.035$, respectively. 
For a given  temperature 
the intensity at the FS increases with increasing doping showing, 
consistently with experiments,\cite{kaminsky05,kim98} a more clearly defined  QP  peak 
with overdoping.
It is also interesting the following behavior. 
For instance, for  $T=0.035$, with decreasing 
doping while the peak at the Fermi level loses intensity a broad shoulder
appears  at positive $\omega$ together with an apparent  gap-like (which is not true gap) structure. 
As suggested by ARPES experiments,\cite{norman1998,norman98} with 
increasing temperature the gap-like  features seem to be filling up but are not closing (see results for $\delta=0.10$). Notice that for $\delta=0.10$  the leading edge below the FS is at $\omega=-0.05t$ which, 
using $t=0.4eV$, is about $20-30 meV$ thus, of the order of  the observed PG value.\cite{norman1998} 
Therefore, spectral functions are strongly temperature dependent, 
even at 
temperatures much lower than the hopping $t$,  showing broad 
features at the antinodal-$k_F$. Spectral functions are nearly doping and temperature independent at the 
nodal-$k_F$ and show well defined QP peaks as in experiments.\cite{kaminsky05,kim98} 
As mentioned above, at mean field level when the  temperature decreases a gap opens at $T^*$ near hot spots.
However, beyond mean field, spectral functions are very incoherent  above $T^*$ 
showing that features expected to occur for $T<T^*$ already appear for   $T>T^*$ due to self-energy renormalizations. 
For instance, the coherence loss when going from nodal-$k_F$ to antinodal-$k_F$ may suggest the appearance of arcs. Therefore, we think that self-energy corrections could be masking a likely  abrupt change at $T^*$ given the appearance, in agreement with observations, 
of a smooth crossover. In addition, it may probably be  that $T^*$ determined from the ARPES line-shape is somewhat shifted from 
the temperature at which PG opens. 

The above  results are due to  the doping and temperature behavior of  the self-energy. 
In Fig.2e  we present results for $Im \Sigma_{flux}$ at $T=0.015$ for several $\delta$ values 
at the antinodal-$k_F$. While for  $\delta=0.20$ $Im \Sigma_{flux}$ shows, as expected for a FL,
the maximum at the Fermi level, with  
decreasing doping (see for instance double-dotted-dashed line for $\delta=0.13$) the maximum is shifted respect to $\omega=0$.
As discussed in Refs.[\onlinecite{katanin03,dellanna06}], this behavior can damage the quasiparticle FL picture. 
For nodal-$k_F$ the situation is FL-like,  {\it e.g.},   $Im \Sigma_{flux}$ shows, for all doping values, the
maximum at $\omega=0$
which increases with $T$ as $\sim T^2$. 
Notice the similarities between our self-energy results in Fig.2e and those discussed in Ref.[\onlinecite{dellanna06}]. 
\begin{figure}
\begin{center}
\setlength{\unitlength}{1cm}
\includegraphics[width=9cm,angle=0.]{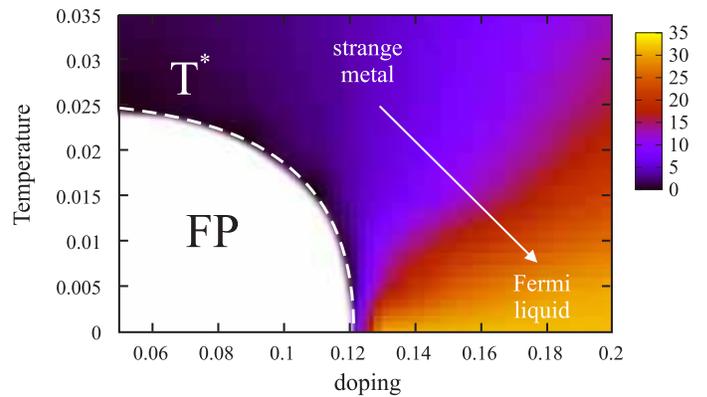}
\end{center}
\caption{(Color online)
$\omega=0$ spectral function intensity at the antinodal-$k_F$ in the $\delta-T$ plane.
}
\label{fig:fase}
\end{figure}

In Fig.3 the $\omega=0$ spectral function intensity  at the  antinodal-$k_F$ is presented in the $\delta$-$T$ plane.
The spectral intensity distribution suggests a crossover from a strange metal at low doping and high temperature to a FL at high doping and low temperature. Based on different arguments, recent  works have 
proposed a similar picture. \cite{castro04}

Before closing let us discuss the influence of $\Sigma_{R\lambda}$ on previous results. In Fig.4a  self-energy results 
for $\Sigma_{R\lambda}$, $\Sigma_{flux}$ and $\Sigma_T$,
at the antinodal-$k_F$, are presented for $\delta=0.13$ and $T=0.02$. 
$Im \Sigma_{R\lambda}$ (solid line) shows a FL-like behavior near $\omega=0$. 
The dotted-dashed line is the result for $Im \Sigma_T$ (eq.(1)). 
While at large frequency $\Sigma_T$ is dominated by $\Sigma_{R\lambda}$, properties near  the FS are dominated by $\Sigma_{flux}$. 
$Im \Sigma_T$ at the nodal-$k_F$ is very similar to $Im \Sigma_{R \lambda}$ (solid line) because 
$\Sigma_{flux}$ has not important influence in the nodal direction.
In Fig.4b, using $\Sigma_T$, the total spectral functions at the nodal-$k_F$ (solid line) and at the antinodal-$k_F$ (dashed line) are presented for low $\omega$. For the spectral function behavior  at large $\omega$ see Ref.[\onlinecite{bejas06}].
Although  $\Sigma_{R \lambda}$ might modify our previous results, it remains 
the main tendency arising from $\Sigma_{flux}$, {\it i.e.},  the spectral functions  are strongly anisotropic on the FS,
losing intensity and becoming broad when going from nodal-$k_F$ to antinodal-$k_F$.
\begin{figure}
\begin{center}
\setlength{\unitlength}{1cm}
\includegraphics[width=7cm,angle=0.]{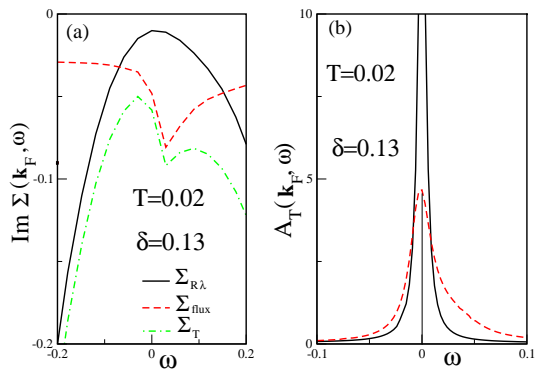}
\end{center}
\caption{(Color online)
(a) Scattering rate at the antinodal-$k_F$ for $\delta=0.13$ and $T=0.02$ for $\Sigma_{flux}$ (dashed line), $\Sigma_{R\lambda}$ (solid line) and 
$\Sigma_T$ (dotted-dashed line). (b) Total spectral function (calculated using $\Sigma_T$) for $\delta=0.13$ and $T=0.02$ at the nodal-$k_F$ (solid line) and antinodal-$k_F$ (dashed line). The intensity at the nodal-$k_F$ was cut for clarity reasons.
}
\label{fig:fase}
\end{figure}

Concluding, in the framework of the $t-J$ model we have computed self-energy corrections in the vicinity of the 
FP instability. 
The corresponding spectral functions 
are anisotropic on the FS and very incoherent near hot spots thus,
they show features similar to those observed in experiments.
In agreement with recent discussions (see Ref.[\onlinecite{chang07}] and references their in) our results show two contributions to the scattering channel. 
One contribution ($\Sigma_{flux}$) is strongly anisotropic on the FS, the other one 
($\Sigma_{R \lambda}$) is very isotropic. 
$\Sigma_{flux}$ dominates at low doping (near $\delta_c$) and low energy 
being relevant  for the PG properties. In contrast, $\Sigma_{R \lambda}$ dominates at large doping 
and large $\omega$. 
In Ref.[\onlinecite{nakamae03}], by means of transport measurements,  it was found that an isotropic  scattering channel dominates in overdoped samples which agrees with our predictions for $\Sigma_{R\lambda}$. 
In a recent work\cite{HEF} it was proposed that the 
high energy features observed in ARPES\cite{exp} may be due to the large $\omega$ behavior of $\Sigma_{R\lambda}$. 
Finally, our results indicate that some problems, which can be interpreted against the FP scenario, may be less serious after including self-energy renormalizations. 

The author thanks to M. Bejas, A. Foussats, H. Parent, A. Muramatsu, and R. Zeyher for valuable discussions.


\begin{thebibliography}{10}

\bibitem{timusk99} T. Timusk and B. Statt, Rep. Prog. Phys. {\bf 62}, 61 (1999).

\bibitem{chakravarty01} S. Chakravarty, R. B. Laughlin, D. K. Morr, and C. Nayak, Phys. Rev. B {\bf 63}, 094503 (2001).

\bibitem{affleck88} I. Affleck and J. B. Marston, Phys. Rev. B {\bf 37}, 3774 (1988).
T. C. Hsu, B. J. Marston, and I. Affleck, Phys. Rev. B {\bf 43}, 28661 (1988)

\bibitem{morse90} D. C. Morse and T. C. Lubensky, Phys. Rev. B {\bf 42}, 7994 (1990).

\bibitem{cappelluti99} E. Cappelluti and R. Zeyher, Phys. Rev. B {\bf 59}, 6475 (1999).

\bibitem{foussats04} A. Foussats and A. Greco,  Phys. Rev. B {\bf 70}, 205123 (2004).

\bibitem{leung00} P. W. Leung, Phys. Rev. B {\bf 43}, 6612 (R) (2000).

\bibitem{chakravarty03} S. Chakravarty, C. Nayak, and S. Tewari, Phys. Rev. B {\bf 68}, 100504 (R) (2003).

\bibitem{norman98}
M. R. Norman, H. Ding, M. Randeria, J. C. Campuzano, T. Yokoya, 
T. Takeuchi, T. Takahashi, T. Mochiku, K. Kadowaki, P. Guptasarma, 
D. G. Hinks,
Nature {\bf 392}, 157 (1998). 

\bibitem{zeyher02} R. Zeyher and A. Greco, Phys. Rev. Lett. {\bf 89}, 177004 (2002).
A. Greco and R. Zeyher,  Phys. Rev. B {\bf 70}, 024518 (2004).

\bibitem{kando07}
T. Kondo, Tsunehiro Takeuchi, Adam Kaminski, Syunsuke Tsuda, and Shik Shin,
Phys. Rev. Lett. {\bf 98}, 267004 (2007).
M. Hashimoto, T. Yoshida, K. Tanaka, A. Fujimori, M. Okusawa, S. Wakimoto, 
K. Yamada, T. Kakeshita, H. Eisaki, and S. Uchida,
Phys. Rev. B {\bf 75}, 140503 (R) (2007).

\bibitem{letacon07}
M. Le Tacon, A. Sacuto, Y. Gallais, D. Colson, and A. Forget,
Phys. Rev. B {\bf 76}, 144505 (2007).

\bibitem{wen07} H.-H. Wen and X.-G. Wen, Physica C {\bf 460}, 28 (2007).

\bibitem{chia07}
Elbert E. M. Chia, Jian-Xin Zhu, D. Talbayev, R. D. Averitt, 
and A. J. Taylor,
Phys. Rev. Lett. {\bf 99}, 147008 (2007).

\bibitem{theo} B. Valenzuela and E. Bascones, Phys. Rev. Lett. {\bf 98}, 227002 (2007). T. Das, R. S. Markiewicz, and A. Bansil, arXiv:0711.0480.

\bibitem{tallon01} J. L. Tallon and J. W. Loran, Physica C {\bf 349}, 53 (2001).

\bibitem{kim98}
C. Kim, P. J. White, Z.-X. Shen, T. Tohyama, Y. Shibata, S. Maekawa, 
B. O. Wells, Y. J. Kim, R. J. Birgeneau, and M. A. Kastner,
Phys. Rev. Lett. {\bf 80}, 4245 (1998).

\bibitem{renner98} 
Ch. Renner, B. Revaz, J.-Y. Genoud, K. Kadowaki, and \O{}. Fischer,
Phys. Rev. Lett. {\bf 80}, 149 (1998).

\bibitem{norman1998} 
C. Norman, M. Randeria, H. Ding,  and J. C. Campuzano,
Phys. Rev. B {\bf 57}, 11093 (R) (1998).

\bibitem{lee06} P. Lee, N. Nagaosa, and X.-G. Wen, Rev. Mod. Phys. {\bf 78}, 17 (2006).

\bibitem{bejas06} M. Bejas, A. Greco, and A. Foussats,  Phys. Rev. B {\bf 73}, 245104 (2006).

\bibitem{model} For simplicity and without losing generality we have considered the minimal $t-J$ model. The inclusion of hopping between second, third, etc  neighbors sites produces only small quantitative but not qualitative changes. 

\bibitem{flux} This form factor is obtained after projecting 
the electron-boson vertex (eq.(9) in Ref.[\onlinecite{foussats04}]) on the corresponding eigenvector for the FP instability.

\bibitem{tp} The inclusion of a hopping between second neighbors $t'/t=0.35$ 
simply shifts $\delta_c$ to values closer to the experimental\cite{tallon01} $\delta_c \sim 0.19$
(see Ref.[\onlinecite{zeyher02}]). 

\bibitem{2x2} $\delta R$ is the fluctuation of the charge Hubbard operator $X^{00}$ 
($X^{00}=\delta(1+\delta R)$) associated to  the number of holes. $\delta \lambda$ is the fluctuation of the Lagrangian 
multiplier related to the nondouble occupancy constraint $X^{00}+\sum_{\sigma} X^{\sigma \sigma}=1$. 

\bibitem{yoshida03}
T. Yoshida, X. J. Zhou, T. Sasagawa, W. L. Yang, P. V. Bogdanov, 
A. Lanzara, Z. Hussain, T. Mizokawa, A. Fujimori, H. Eisaki, 
Z.-X. Shen, T. Kakeshita, and S. Uchida,
Phys. Rev. Lett. {\bf 91}, 027001 (2003).

\bibitem{kaminsky05}
A. Kaminski, H. M. Fretwell, M. R. Norman, M. Randeria, S. Rosenkranz, 
U. Chatterjee, J. C. Campuzano, J. Mesot, T. Sato,
T. Takahashi, T. Terashima, M. Takano, K. Kadowaki, Z. Z. Li, and H. Raffy,
Phys. Rev. B {\bf 71}, 014517 (2005).

\bibitem{chang07}
J. Chang, M. Shi, S. Pailhes, M. Maansson, T. Claesson, O. Tjernberg, 
A. Bendounan, L. Patthey, N. Momono, M. Oda, M. Ido, C. Mudry, J. Mesot,
arXiv:0708.2782.

\bibitem{katanin03} A. A. Katanin and A. P. Kampf, Phys. Rev. Lett. {\bf 93}, 106406 (2004).

\bibitem{dellanna06} L. Dell'Anna and W. Metzner, Phys. Rev. B {\bf 73}, 045127 (2006).

\bibitem{castro04} H. Castro and G. Deutscher,  Phys. Rev. B {\bf 70}, 174511 (2004).
A. Kaminski, S. Rosenkranz, H. M. Fretwell, Z. Z. Li, H. Raffy, 
M. Randeria, M. R. Norman, and J. C. Campuzano,
Phys. Rev. Lett. {\bf 90}, 207003 (2003).

\bibitem{nakamae03}
S. Nakamae, K. Behnia, N. Mangkorntong, M. Nohara, H. Takagi, 
S. J. C. Yates, and N. E. Hussey,
Phys. Rev. B {\bf 68}, 100502 (R) (2003).

\bibitem{HEF} A. Greco, Solid State Communications {\bf 142}, 318 (2007).

\bibitem{exp}
B. Xie, K. Yang, D. W. Shen, J. F. Zhao, H. W. Ou, J. Wei, S. Y. Gu, 
M. Arita, S. Qiao, H. Namatame, M. Taniguchi, N. Kaneko, H. Eisaki, 
K. D. Tsuei, C. M. Cheng, I. Vobornik, J. Fujii,5 G. Rossi, Z. Q. Yang, 
and D. L. Feng
Phys. Rev. Lett. {\bf 98}, 147001 (2007).
Z.-H. Pan, P. Richard, A.V. Fedorov, T. Kondo, T. Takeuchi, S.L. Li, 
Pengcheng Dai, G.D. Gu, W. Ku, Z. Wang, H. Ding
cond-mat/0610442.


\end{thebibliography}
\end{document}